\documentclass[twocolumn,aps, PRL, twocolumn]{revtex4}
\usepackage[T1]{fontenc}
\usepackage[latin1]{inputenc}

\makeatletter

\providecommand{\LyX}{L\kern-.1667em\lower.25em\hbox{Y}\kern-.125emX\@}

\makeatother

\begin{document}

\title{Comment on: Experimental Simulation of Two-Particle Quantum Entanglement using
Classical Fields}

\author{A. F. Kracklauer}

\address{kracklau@fossi.uni-weimar.de}

\maketitle
In a recent Letter Lee and Thomas (LT) report on an experiment technique that
enabled them to \emph{simulation} (emphasis added) ``a four particle entangled
state''\cite{LT} Their results conform precisely with those seen is similar
experiments in which the observed correlations are those obtained in demonstrations
of GHZ 4-fold `photon' experiments, as reported elsewhere; e.g.:\cite{Pan}
These latter experiments are taken to demonstrate the quantum mechanical nature
of EPR correlations, in particular the multi-particle GHZ correlations. The
observation of the identical correlations among classical fields evokes the
question: how can a ``quantum'' phenomena be rendered classically? This question
is particularly acute in view of the conclusion of Bell's ``Theorem,'' which
states that such correlations are impossible in ``local, realistic,'' circumstances,
i.e., for \emph{all} classical phenomena.\cite{Bell} 

The premise of this comment is based on the proposition that nature does not
\emph{``simulate''} itself, and that the conceptual implications of LT's experiment
are therefore that it is an empirical counterexample to Bell's Theorem, rigorously
establishing its invalidity.

While this observation is in sharp contrast with much prevailing opinion regarding
the significance of Bell's analysis, in fact it is supported by an ever increasing
body of research which in various ways also leads to the conclusion that there
is a fatal error in Bell's Theorem. This critical research falls into three
categories: 1.) the logical deconstruction of Bell and Bell-Kochen-Specker Theorems,
2.) the construction of local realistic models and simulations of EPR correlations,
i.e., theoretical counterexamples, and 3.) empirical counterexamples, of which
LT's is only the most recent. Additionally, there is much similar research which
achieves only partial or specialized results, e.g., \cite{FPP}. There is, however,
only fragmented agreement on the logical interrelationships among these results.
None of it, in any case, has been seriously refuted although all of it is oft
rejected on the basis of conservative antipathy.

Apparently the first category was initiated by Jaynes.\cite{Jaynes} He argued
that Bell's fundamental premise, namely that ``locality'' demands that the coincident
correlation for EPR situations factor according to: \( P(a,\, b,\, \lambda )=P(a\, |\lambda )P(b\, |\lambda )P(\lambda ) \),
is a misconstrual of Bayes' formula, namely: \( P(a,\, b,\, \lambda )=P(a\, |b,\, \lambda )P(b\, |\lambda )P(\lambda ) \).\cite{Feller}
Bayes's formula does not imply instantaneous communication between the two space-like
separated detectors as it also accommodates ``common-cause'' correlation. This
result has been rediscovered independently in various styles of argument at
least five times.\cite{Per}-\cite{LBL} Similarly, Barut seems to have been
first to find the lacuna in Kochen-Specker type, inequality-free proofs of Bell's
Theorem.\cite{Barut} The error there consists in failing to discern which spin
operators are physically relevant \emph{at the same time}. See also \cite{mepris}. 

Model-making turns out to be rather simple so long as no effort is made to simultaneously
satisfy Bell's factorization. Utterly straightforward models exist for all the
standard EPR-B (EPR-Bohm; i.e., polarization entangled) experiments.\cite{Entang}
Numerical, ``Monte Carlo,'' simulations devoid of `quantum structure' for 2-fold
EPR experiments also have violated Bell Inequalities.\cite{WH}

In addition to LT's 4-fold experiment, Evdokimov et al. reported on a 2-fold
EPR ``simulation.''\cite{Evdo} Again, the terminology employed respects sensibilities
at the expense of precision; in fact any classical phenomena that exhibit EPR
correlations, contradict Bell.

Finally, note that rejecting the concept of ``quantum nonlocal correlation''
has scant effect on the discipline of physics except to expunge paradoxical
jargon. However, the larger issue brought up by EPR, i.e., the search for a
deeper theory completing Quantum Mechanics, in accord with these conclusions,
can not be considered a quixotic endeavor.

\end{document}